\documentclass[fp,twocolumn]{jpsj3}
\usepackage{txfonts}

\usepackage{color}

\title{Machine Learning Quantum States -- Extensions to Fermion-Boson Coupled Systems and Excited-State Calculations}

\author{Yusuke Nomura$^1$\thanks{yusuke.nomura@riken.jp}}
\inst{$^1$RIKEN Center for Emergent Matter Science, 2-1 Hirosawa, Wako, 351-0198, Japan} 

\abst{
To analyze quantum many-body Hamiltonians, recently, machine learning techniques have been shown to be quite useful and powerful. 
However, the applicability of such machine learning solvers is still limited. 
Here, we propose schemes that make it possible to apply machine learning techniques to analyze fermion-boson coupled Hamiltonians and to calculate excited states. 
As for the extension to fermion-boson coupled systems, we study the Holstein model as a representative of the fermion-boson coupled Hamiltonians. 
We show that the machine-learning solver achieves highly accurate ground-state energy, improving the accuracy substantially compared to that obtained by the variational Monte Carlo method.
As for the calculations of excited states, we propose a different approach than that proposed in K. Choo {\it et al.}, Phys. Rev. Lett. {\bf 121} (2018) 167204.
We discuss the difference in detail and compare the accuracy of two methods using the one-dimensional $S=1/2$ Heisenberg chain. 
We also show the benchmark for the frustrated two-dimensional $S=1/2$ $J_1$-$J_2$ Heisenberg model and show an excellent agreement with the results obtained by the exact diagonalization.  
The extensions shown here open a way to analyze general quantum many-body problems using machine learning techniques. 
}

\begin{document}
\maketitle

\section{Introduction}

Solving quantum many-body Hamiltonians with high accuracy is a great challenge in condensed-matter physics. 
The behaviors of particles in such quantum systems are governed by many-body wave functions; 
once exact many-body wave functions are given, we can predict the properties of many-body systems.  
Therefore, a goal is to obtain exact many-body wave functions. 
However, in general, many-body wave functions are vectors with huge dimensions growing exponentially with the number of particles, which makes exact diagonalization intractable for large systems. 
Given this situation, it is imperative to represent many-body wave functions accurately with a computationally tractable number of parameters. 

For this problem, machine learning techniques can play a role. 
Machine learning is powerful in extracting essential features from big data. 
Therefore, it would also be useful in extracting the essential patterns of the many-body wave functions and obtaining compact representations. 

Indeed, Carleo and Troyer~\cite{Carleo602} have introduced variational ansatz for many-body ground states based on the restricted Boltzmann machine (RBM). 
The RBM is a kind of artificial neural network~\cite{RBM_Smolensky} consisting of visible and hidden units and is used to approximate probability distribution over the visible unit configurations. 
If we interpret the values of many-body wave functions as a generalized probability (allowing complex numbers) and make a mapping between physical and visible unit configurations,  
we can represent many-body wave functions in terms of RBM. 
The RBM wave functions were first applied to spin systems and have been shown to be able to represent ground states of spin Hamiltonians in high accuracy.~\cite{Carleo602} 

One of the advantages of using the RBM is its flexible representative power.   
The RBM can represent quantum states showing volume-law entanglement entropy.~\cite{PhysRevX.7.021021,PhysRevB.97.085104} 
If an arbitrarily large number of hidden units are introduced, it can represent any bounded continuous function to arbitrary accuracy (universal approximation).~\cite{doi:10.1162/neco.2008.04-07-510,doi:10.1162/NECOa00113}
Another advantage is that the RBM allows us to approximate quantum many-body states in an unbiased way. 
This is in contrast to the conventional wave function methods where most calculations assume a specific form of the wave function, i.e., the calculations are biased. 

After Ref.~\citen{Carleo602}, various studies have shown the usefulness of using machine learning techniques to study quantum many-body systems. 
Several studies have introduced different neural network than the RBM~\cite{Gao_Ncom,PhysRevB.99.165123,DBM_NC,doi:10.1142/S0219749918400087,doi:10.7566/JPSJ.86.093001,doi:10.7566/JPSJ.87.014001,doi:10.7566/JPSJ.87.074002,PhysRevB.97.035116,PhysRevB.100.125124,Westerhout_2020}. 
The relationship between the RBM and tensor network methods has also been discussed.~\cite{PhysRevB.97.085104,PhysRevX.8.011006,1751-8121-51-13-135301,Huang_arXiv}
The applicability of machine learning solvers has been extended to, for example, 
frustrated spin systems~\cite{PhysRevB.97.035116,PhysRevB.98.104426,PhysRevB.100.125124,PhysRevB.100.125131,Westerhout_2020},
itinerant boson systems~\cite{doi:10.7566/JPSJ.86.093001,doi:10.7566/JPSJ.87.014001},
topological states~\cite{PhysRevX.7.021021,PhysRevB.96.195145,PhysRevX.8.011006,1751-8121-51-13-135301,PhysRevB.99.155136,PhysRevB.97.195136,Huang_arXiv}, 
fermion systems~\cite{PhysRevB.97.035116,PhysRevB.96.205152,PhysRevLett.122.226401,Han_arXiv,Choo_fermion_arXiv}, 
excited states~\cite{PhysRevLett.121.167204,Vieijra_2020},
open quantum states~\cite{PhysRevLett.122.250501,PhysRevLett.122.250502,PhysRevLett.122.250503,PhysRevB.99.214306}, 
and 
quantum states with nonabelian or anyonic symmetries~\cite{Vieijra_2020}.

In the present paper, we report two crucial extensions of machine learning solvers. 
One is an extension to the fermion-boson coupled systems. 
The other is the calculations of excited states. 
For the fermion-boson coupled systems, we select electron-phonon coupled systems as an example. 
A generalization of the proposed scheme to other fermion-boson systems is straightforward. 
As for the calculations of excited states, we propose a different approach than that employed in Ref.~\citen{PhysRevLett.121.167204}.

This paper is organized as follows. 
Secs.~\ref{Sec:el_ph} and \ref{sec_excited} are devoted to the extension to fermion-boson coupled systems and calculations of excited states, respectively. 
In both sections, we provide the details of the methods and demonstrate the accuracy of the machine learning methods.
We show the summary and future perspective in Sec.~\ref{sec_summary}.

\section{Extension to electron-phonon coupled systems}
\label{Sec:el_ph}

The electron-phonon coupling is a fundamental interaction in solids. 
In the conventional superconductors, the electron-phonon coupling gives a ``glue" for creating Cooper pairs.
Even in the strongly-correlated materials, the coupling to the phonons plays an important role. 
For example, in transition metal oxides, the metal-insulator transitions often accompany structure transition~\cite{RevModPhys.70.1039}. 
In the alkali-doped fullerides~\cite{RevModPhys.69.575,RevModPhys.81.943}, unusual cooperation between strong correlation and phonons induces unconventional $s$-wave superconductivity next to the Mott insulating phase~\cite{Nomurae1500568,Nomura_2016}. 
The coupling to the optical phonons of SrTiO$_3$ substrate has been suggested to be the origin of enhanced superconductivity in FeSe thin film on the SrTiO$_3$ substrate~\cite{Lee_Nature}.

Therefore, in order to get a deeper understanding of electron-phonon physics, it is essential to develop powerful numerical methods to analyze electron-phonon coupled Hamiltonians. 
Here, we discuss an extension of the machine-learning wave function methods to the electron-phonon coupled systems.

\subsection{Model}
Hereafter, as a representative example, 
we restrict ourselves to the case of the one-dimensional spinless Holstein model and discuss how we construct machine-learning wave functions. 
The Hamiltonian is given by 
\begin{eqnarray}
{\mathcal H }  \!  =  \!  -t \! \sum_{i}  (\hat{c}^\dagger_i \hat{c}_{i+1} \!   + \! {\rm h.c.} )   - \!  g \! \sum_i  \bigl ( \hat{n}_i - \frac{1}{2} \bigr )  \bigl (\hat{b}_i^\dagger + \hat{b}_i \bigr)  + \!    \!  \sum_i  \! \omega  \hat{b}^\dagger_i \hat{b}_i  , 
\label{Eq.1D_H_model}
\end{eqnarray}
where $\hat{c}^\dagger_i$ ($\hat{c}_i$) is the creation (annihilation) operator for the electron at site $i$ and  $ \hat{b}^\dagger_i$ ($ \hat{b}_i$) creates (annihilates) phonon at site $i$.  
$t$ is a hopping amplitude, $g$ is the strength of the electron-phonon coupling ($g>0$), and $\omega$ is the phonon frequency. 
$\hat{n}_i = \hat{c}^\dagger_i \hat{c}_i $ is the number operator. 
The dimensionless displacement operator of the phonons is given by $\hat{x}_i  = \frac{1} { \sqrt{2 } } ( \hat{b}_i^\dagger + \hat{b}_i ) $.
For simplicity, we take the mass of nuclei to be 1. 
The methods presented here can be easily extended to more general electron-phonon coupled systems and also even more general fermion-boson coupled systems. 

For later use, it is convenient to define occupation and dimensionless displacement operators ($\hat{n}^{\rm s}_i$ and $\hat{x}^{\rm s}_i $, respectively) in a staggered way: 
\begin{eqnarray}
\hat{n}^{\rm s}_i  =    (-1)^i ( 2 \hat{n}_i - 1 )  
\end{eqnarray}
and 
\begin{eqnarray}
\hat{x }^{\rm s}_i  =   (-1)^i  \hat{x}_i  =   \frac{ (-1)^i  } {\sqrt{2}  }   \bigl (\hat{b}_i^\dagger + \hat{b}_i \bigr) . 
\end{eqnarray}
Physically, $\hat{n}^{\rm s}_i$ is proportional to the electron and hole occupations (with constant shift) for even and odd sites, respectively.  
Introduction of $\hat{x}^{\rm s}_i $ corresponds to the change of positive direction of phonon displacements depending on whether the site index is even or odd. 
With this definition, the instability for the charge-density wave with the momentum $\pi$ can be interpreted as the ferroic order in the $n^{\rm s}_i$ occupation. 
The Hamiltonian is rewritten as 
\begin{eqnarray}
\label{Eq_H_Holstein}
{\mathcal H } = -t \sum_{i}  (\hat{c}^\dagger_i \hat{c}_{i+1}  +{\rm h.c.} )   - \frac{g}{ \sqrt{2}}   \sum_i   \hat{n}^{\rm s}_i   \hat{x}^{\rm s}_i  + \sum_i \omega  \hat{b}^\dagger_i \hat{b}_i.
\end{eqnarray}

As for the basis, we use the occupation number basis of electrons and phonons in real space; it is a natural choice because the Hamiltonian in Eq. (\ref{Eq_H_Holstein}) is defined in real space. 
The electron occupations are specified by $n^{\rm s}_i =  (-1)^i ( 2 n_i - 1 )  $.
The phonon basis set $ \{ | m_i \rangle    \} $ with the phonon occupation number $m_i$ ($\geq 0$)  is defined such that all the matrix elements of the $\hat{x}^{\rm s}_i$ ($ = (-1)^i \hat{x}_i $) operator become positive.
More specifically, we define $ | m_i \rangle $ as $ | m_i \rangle  =  \left( (-1)^{i} \right)^{m_i } | m_i \rangle_0 $  with $| m_i \rangle_0$ being the standard definition of phonon occupation state.
Then, the matrix elements become positive as $ \langle m_i  |  \hat{x}^{\rm s}_i   | m_i +1 \rangle  = \sqrt{ \frac{ m_i +1 }{2} } $. 


\subsection{Methods}

\subsubsection{Recent variational wave functions} 

Recently, there have been several proposals for the variational wave functions for electron-phonon coupled systems~\cite{PhysRevB.89.195139,PhysRevB.96.205145}. 
In those studies, the form of the variational wave functions reads 
\begin{eqnarray}
  | \Psi \rangle   =     P_{\rm el \mathchar`- ph} \   | \Psi_{\rm el}  \rangle \otimes   | \Psi_{\rm ph}  \rangle.  
\end{eqnarray}
Here,  $| \Psi_{\rm el} \rangle$ and $| \Psi_{\rm ph}  \rangle$ are the electron and phonon wave functions, respectively. 
$P_{\rm el \mathchar`- ph}$ is the electron-phonon correlation factor. 
In this form, the correlations between electrons and phonons are taken into account only by $P_{\rm el \mathchar`- ph}$; without  $P_{\rm el \mathchar`- ph}$, electron and phonon degrees of freedom are decoupled. 

\subsubsection{Machine-learning variational wave function} 

We improve the accuracy of the variational wave function by 
i) preparing electron-phonon-entangled wave function $| \Psi_{\rm el \mathchar`- ph}  \rangle$ even without electron-phonon correlation factor 
and 
ii) improving electron-phonon correlation factor with RBM (denoted as $P^{\rm RBM}_{\rm el \mathchar`- ph}$). 
Then, the form of the variational wave function is given by 
\begin{eqnarray}
\label{Eq.form_wavefunction}
  | \Psi \rangle   =     P^{\rm RBM}_{\rm el \mathchar`- ph} \   | \Psi_{\rm el \mathchar`- ph}  \rangle.  
\end{eqnarray}
In the following, we discuss these two points in more detail. 
\\

\paragraph {i) Electron-phonon-entangled wave function $\mathbf {| \Psi_{\rm \bf el \mathchar`- ph}  \rangle}$.}

As can be seen in Eq. (\ref{Eq_H_Holstein}), when the $n^{\rm s}_i$ occupation is positive $n^{\rm s}_i \! = \! 1$ (negative $n^{\rm s}_i \! = \! -1$), the positive (negative)  $x^{\rm s}_i$ displacement is energetically favored.  
As we will show in the following, even without the electron-phonon correlation factor, one can take into account this primitive correlation between electrons and phonons.

Defining real-space electron and phonon configurations as $\nu =  (  n^{\rm s}_1, n^{\rm s}_2, ..., n^{\rm s}_{N_{\rm site}} )$ and $\mu = (  m_1, m_2, ..., m_{N_{\rm site}} )$, respectively, the electron-phonon-entangled wave function reads 
\begin{eqnarray}
\label{Eq_Psi_el-ph}
| \Psi_{\rm el \mathchar`- ph}  \rangle =  \sum_{\nu , \mu } | \nu ,\mu \rangle   \Psi_{\rm el} (\nu) \Psi_{\rm ph} (\mu ; \nu), 
\end{eqnarray}
where the phonon wave function $\Psi_{\rm ph}$ for the phonon configuration $\mu$ does depend on the electron configuration $\nu$. 
Note that, in the previous studies,~\cite{PhysRevB.89.195139,PhysRevB.96.205145} the phonon wave function $\Psi_{\rm ph}$ is independent of the electron configuration $\nu$. 

The phonon part, which depends on the electron configuration $\nu$, is given by
 \begin{eqnarray}
\left  |  \Psi_{\rm ph} (\nu)  \right \rangle =  \prod_i  \left (   \sum_{m_i=0}^{m_{\rm max} } c_{m_i} (n^{\rm s}_i) | m_i \rangle  \right ). 
\label{Eq.form_phonon}
 \end{eqnarray}
Here, $m_{\rm max}$ is the maximum phonon occupation number and the coefficient $c_{m_i} (n^{\rm s}_i)$ is a variational parameter. 
In this case, the coefficients for the phonon states depend on local electron configuration, which allows us to take into account the above described primitive local correlation between electrons and phonons. 

\begin{figure}[t]
\begin{center}
\includegraphics[width=0.42\textwidth]{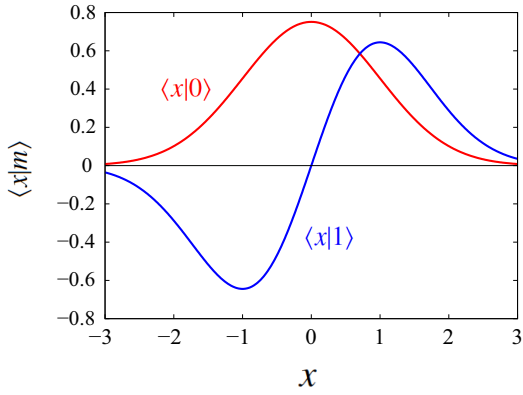}
\caption{ 
(Color online)
Phonon wave functions  $  \langle x | 0 \rangle$   and  $  \langle x | 1 \rangle $.  
$x$ is the dimensionless displacement.
$  \langle x | m \rangle $ is given by $  \langle x | m \rangle   = ( \! \sqrt{\pi} 2^m  m ! )^{-1/2} H_m( x ) \exp( -x^2/2)$, where $H_m( x )$ is the Hermite polynomial.
Here, we omit site index $i$. 
}
\label{fig_phonon_basis}
\end{center}
\end{figure}

If the $c_{m_i}$ coefficients have the same sign for all occupation numbers $m_i$, the expectation value of the $x^{\rm s}_i $ displacement takes positive value. 
This can be easily understood considering the fact that all the matrix elements $ \langle m_i  |  \hat{x}^{\rm s}_i   |m_i +1 \rangle $ are positive (see also Fig.~\ref{fig_phonon_basis}). 
This situation is favored when the $n^{\rm s}_i$ occupation is positive ($n^{\rm s}_i \! = \! 1 $). 
On the other hand, when $n^{\rm s}_i \!=  \! - 1 $, negative $x^{\rm s}_i $ will be induced.
The negative $x^{\rm s}_i $ is realized if the $c_{m_i}$ coefficients have the sign change between even and odd occupation numbers. 
To satisfy the above described primitive correlation, for $n^{\rm s}_i \! = \! 1 $, 
we initialize $c_{m_i}$ coefficients such that $c_{m_i} (n^{\rm s}_i \! = \! 1 ) > 0 $ for all $m_i$.
On the other hand, the $c_{m_i}$ coefficients for $n^{\rm s}_i  \! = \! - 1$ are initialized as $c_{m_i} (n^{\rm s}_i \! = \! - 1) > 0$ and $c_{m_i} (n^{\rm s}_i  \! = \! - 1) < 0$  for even and odd $m_i$, respectively.  
This initialization gives the energy gain in the electron-phonon coupling term $\bigl( -\frac{g}{ \sqrt{2}}   \sum_i   \hat{n}^{\rm s}_i   \hat{x}^{\rm s}_i \bigr)$ in Eq. (\ref{Eq_H_Holstein}).

We have two independent variational parameters $c_{m_i} (n^{\rm s}_i \! = \! \pm 1 )$ for each phonon occupation $m_i$ ($0 \leq m_i \leq m_{\rm max}$). 
Therefore, the number of variational parameters for the coefficient $c_{m_i} (n^{\rm s}_i)$ is $2 (m_{\rm max} \! + \! 1) N_{\rm site} $. 
In the actual calculations, we impose the translational symmetry, and the number is reduced to $2 (m_{\rm max} \! +\! 1)$.
In the particle-hole symmetric case without symmetry breaking, we can further reduce the number to $m_{\rm max} \! +\! 1$: 
We can set  $c_{2k} (n^{\rm s}_i \!  =\!  1) =  c_{2k} (n^{\rm s}_i \!  =\!  -1 )$ and $c_{2k + 1 } (n^{\rm s}_i  \! =  \! 1) = -  c_{2k + 1 } (  n^{\rm s}_i \! = \! -1 )$ with some non-negative integer $k$.

As for the electron wave function $\Psi_{\rm el}$, we employ the pair-product (geminal) wave function. 
The pair-product wave function has flexible representability: it can describe, for example, Fermi-sea, antiferromagnetic, charge-ordered, and superconducting states.   
The pair-product wave function is given by 
\begin{eqnarray}
| \Psi_{\rm el }  \rangle =  \left( \sum_{i,j}  f_{ij} \hat{c}^\dagger_{i \uparrow}  \hat{c}^\dagger_{j\downarrow}   \right ) ^ { \frac{N_{\rm el}}{2} } |  0 \rangle 
\label{Eq.form_PP}
\end{eqnarray}
with the number of electrons $N_{\rm el}$ and the variational parameter $f_{ij}$.
Here, we write the electron spin degrees of freedom explicitly to make it possible to apply our wave functions to the Hamiltonian with spin degrees of freedom. 
In the case of the present spinless case, we just use one spin component out of up and down components. 
When the boundary condition of the lattice is periodic or anti-periodic, we can take $N_{\rm u.c.}$-site ($N_{\rm u.c.} \leq N_{\rm site} $) unit cell structure for $f_{ij}$ parameters ($f_{i + p N_{\rm u.c.} , j + p N_{\rm u.c.}} = f_{ij}$ for some integer $p$) to reduce the number of variational parameters. 
With this setting, we can study the ordered state whose period is smaller than or equal to $N_{\rm u.c.}$. 
The number of $f_{ij}$ is reduced from $N_{\rm site}^2$ to  $N_{\rm u.c.}  N_{\rm site}$.

\paragraph {ii) RBM electron-phonon correlation factor  $\mathbf P^{\rm \bf RBM}_{\rm \bf el \mathchar`- ph}$.}
In Refs. \citen{PhysRevB.89.195139} and \citen{PhysRevB.96.205145}, the form of the electron-phonon correlation factor reads
\begin{eqnarray}
\label{Eq.P_elph1}
P_{\rm el \mathchar`- ph}  =  \exp \left ( \sum_{ij} \alpha_{ij} n_i x_j \right ) 
\end{eqnarray}
and 
\begin{eqnarray}
\label{Eq.P_elph2}
P_{\rm el \mathchar`- ph}  =  \exp \left ( - \sum_{ij} v_{ij} n_i m_j \right ),  
\end{eqnarray}
respectively ($\alpha_{ij}$ and $v_{ij}$ are variational parameters). 
Both of them are constructed based on physical insight. 
In the present study, we replace them with the RBM correlation factor, which is more flexible and unbiased.  
Indeed, as is shown in Refs.~\citen{1751-8121-51-13-135301} and \citen{PhysRevB.96.205152}, the two-body correlation factors such as Eqs.~(\ref{Eq.P_elph1}) and  (\ref{Eq.P_elph2}) can be analytically expressed by the RBM correlation factor. 
On top of such two-body correlations, the RBM can describe many-body correlations such as three-body and four-body correlations simultaneously in an unbiased way.  
Furthermore, as is discussed in the introduction, it is ensured that any correlations can be expressed exactly by the RBM in the limit of an infinite number of hidden units (``universal approximation").

\begin{figure}[t]
\begin{center}
\includegraphics[width=0.42\textwidth]{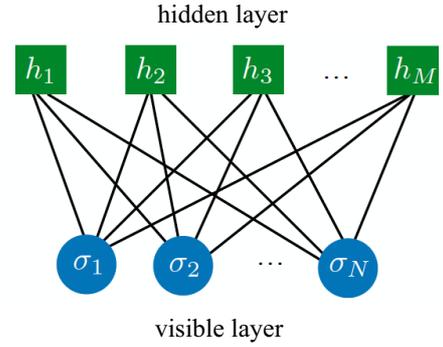}
\caption{
(Color online)
Structure of RBM with $N_{\rm hidden} =M$ and  $N_{\rm visible} =N$.
 $N_{\rm hidden}$ and $N_{\rm visible}$ are the number of hidden and visible units, respectively. 
}
\label{fig_structure_RBM}
\end{center}
\end{figure}

The structure of the RBM is shown in Fig.~\ref{fig_structure_RBM}. 
By identifying visible units configuration $\{ \sigma_l \}$ and that of physical degrees of freedom $(\nu, \mu)$ (see the following), the RBM correlation factor is given by 
\begin{eqnarray}
 P^{\rm RBM}_{\rm el \mathchar`- ph} (\sigma) \! \! \!   \! &=& \! \! \! \!   \sum_{ \{ h_k\}}  \exp \left (   \sum_{l} a_l \sigma_l  + \sum_{l,k} W_{lk}  \sigma_l h_k +\sum_{k} b_k h_k \right ) \nonumber \\
  &= & \! \! \!  \! \exp \left (\sum_{l} a_l \sigma_l \right)  \times  \prod_k   2 \cosh \left( b_k + \sum_{l} W_{lk}  \sigma_l  \right )  \nonumber \\
 \label{Eq.form_RBM}
\end{eqnarray}
Here, $\sigma_l = \pm 1$ and $h_k = \pm 1 $ denote the state of visible and hidden units respectively. 
$\sigma = (\sigma_1, \sigma_2, ..., \sigma_{N_{\rm visible}})$ is the spin configuration of the visible units. 
$\{ a_l, W_{lk}, b_k \}$ are variational parameters. 
We neglect irrelevant one-body $a_l$ terms and optimize only $\{ W_{lk}, b_k \}$ parameters. 

\begin{figure}[t]
\begin{center}
\includegraphics[width=0.42\textwidth]{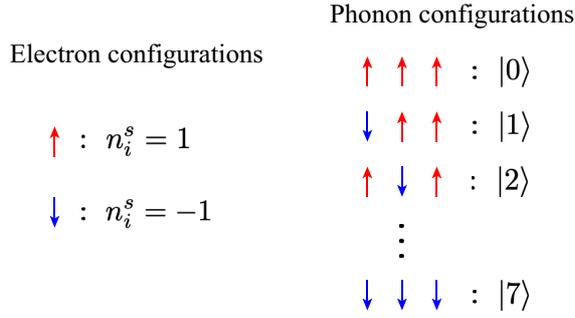}
\caption{
(Color online)
Mapping between visible unit ($\sigma$) configurations and electron and phonon configurations at each site. 
$\uparrow$ and $\downarrow$ correspond to $\sigma_l =1$ and $-1$ with $l$ being the visible-unit index, respectively. 
In this example, three visible units are introduced to represent phonon configurations.
}
\label{fig_definition_sigma}
\end{center}
\end{figure}

In the case of electron-phonon coupled Hamiltonian in Eq. (\ref{Eq_H_Holstein}), to represent both the electron and phonon configurations, 
we prepare one visible unit for electron configuration and $M$ visible units for phonon configuration for each site (see Fig.~\ref{fig_definition_sigma}). 
In total, the number of visible units becomes $N_{\rm visible} = N_{\rm site} (M+1)$.  
We define $\sigma_l =  n^{\rm s}_i = \pm1 $ for the visible units in charge of electron configurations. 
For phonon configuration, with $M$ visible units, we can describe the phonon states up to the occupation number of $2^M -1$ at each site. 
For example, if we have three visible units, we can map $(\sigma_{l_1}, \sigma_{l_2}, \sigma_{l_3}) = (1,1,1)$ onto $| 0 \rangle$, 
$ (-1,1,1)$ onto $| 1 \rangle$,  $ (1,-1,1)$ onto $| 2 \rangle$, ..., and $ (-1,-1,-1)$ onto $| 7 \rangle$.
Because the maximum phonon occupation number $m_{\rm max}$ scales exponentially with $M$, we can essentially simulate infinite $m_{\rm max}$. 
This definition gives one to one correspondence between $ \sigma$ spin configuration and electron and phonon configuration $(\nu, \mu)$. 

The correlation factor $ P^{\rm RBM}_{\rm el \mathchar`- ph} (\sigma) $ in Eq.~(\ref{Eq.form_RBM}) is combined with the electron-phonon-entangled wave function $| \Psi_{\rm el \mathchar`- ph}  \rangle $ [Eq. (\ref{Eq.form_wavefunction})]. 
In addition to the primitive correlation between electrons and phonons taken into account by $|\Psi_{\rm el \mathchar`- ph}  \rangle $, the RBM correlation factor $ P^{\rm RBM}_{\rm el \mathchar`- ph}$ takes into account more sophisticated correlations. 
Note that the RBM correlation factor takes into account not only electron-phonon correlations but also electron-electron and phonon-phonon correlations. 
Therefore, we do not need to introduce additional correlation factor for electron-electron and phonon-phonon correlations, 
whereas the wave functions employed in the previous studies~\cite{PhysRevB.89.195139,PhysRevB.96.205145} use electron-electron correlation factor separately with electron-phonon correlation factor in Eqs. (\ref{Eq.P_elph1}) and (\ref{Eq.P_elph2}). 

The accuracy of the wave function is controlled by the number of hidden units: The more hidden units we introduce, the more accurate the wave function becomes. 
We define the control parameter of accuracy $\alpha$ to be $\alpha = N_{\rm hidden} / N_{\rm visible}$ with the number of hidden unit $N_{\rm hidden}$.
We call $\alpha$ hidden variable density. 
By taking $\alpha$ to be integer and imposing translational symmetry in $W_{ik}$ and $b_{k}$ parameters~\cite{Carleo602}, 
the number of $W_{ik}$ and $b_{k}$ parameters becomes $\alpha N_{\rm visible}$ and $\alpha$, respectively. 

\subsubsection{Optimization}

We employ the stochastic reconfiguration (SR) method~\cite{PhysRevB.64.024512} to optimize the variational parameters~\cite{note_SR}.  
The optimization is done to minimize the energy expectation value $E = \langle \Psi  | {\mathcal H} | \Psi \rangle / \langle \Psi  | \Psi \rangle $, where the expectation value is calculated using the Monte Carlo method over the real-space configurations $(\mu, \nu)$ with the weight $| \Psi (\mu, \nu)| ^2$. 
The SR method realizes imaginary time Hamiltonian evolution within the representative power of the variational wave function~\cite{PhysRevLett.107.070601} and hence enables stable optimization.  
The variational parameters to be optimized are $\{  c_{m_i}(n^{\rm s}_i), f_{ij}, W_{lk}, b_k  \}$ in Eqs.~(\ref{Eq.form_phonon}), (\ref{Eq.form_PP}), and  (\ref{Eq.form_RBM}). 

The numbers of independent parameters for the translationally invariant systems are $2 (m_{\rm max} \! +\! 1)$ for $c_{m_i}(n^{\rm s}_i)$, $N_{\rm u.c.} N_{\rm site}$ for $f_{ij}$, $\alpha N_{\rm visible}$ for $W_{lk}$, and $\alpha$ for $b_{k}$ parameters. 
If the system has particle-hole symmetry, the number for $c_{m_i}(n^{\rm s}_i)$ parameters can be reduced to $m_{\rm max} \! +\! 1$.
See the previous sections for more detail on the number of independent parameters.

\subsection{Results} 

\subsubsection{Total energy} 
To show the accuracy of our wave function, we first show the results for the total energy. 
We consider the one-dimensional spinless Holstein model whose Hamiltonian is given by Eq. (\ref{Eq_H_Holstein}).
Following the previous studies~\cite{PhysRevB.53.9676,PhysRevB.89.195139}, 
we use periodic (anti-periodic) boundary condition when the number of electrons is odd (even). 
We simulate 6-, 8-, and 16-site systems with $\omega/t = 1$, $g/t = 1.5$. 
The filling is set to be half-filling. 
In this setting, the ground state is described by Tomonaga-Luttinger liquid (TLL).
Because the system has particle-hole symmetry, the number of independent phonon coefficients  $\{ c_{m_i}(n_i) \}$ becomes $m_{\rm max} \! +\! 1$ (see Methods section). 
We set maximum phonon occupation number $m_{\rm max}$ at each site to be 15, i.e., the number for visible units for phonon configuration is $M = 4$ per site.
Then the total number of visible units are  $N_{\rm visible} = 5N_{\rm site}$. 
We confirm the convergence of the results with respect to $m_{\rm max}$. 
As for the $f_{ij}$ parameters, we take 6-site unit-cell structure for the 6-site system and 4-site unit-cell structure for the 8-site and 16-site systems.

\begin{figure*}[t]
\begin{center}
\includegraphics[width=0.9\textwidth]{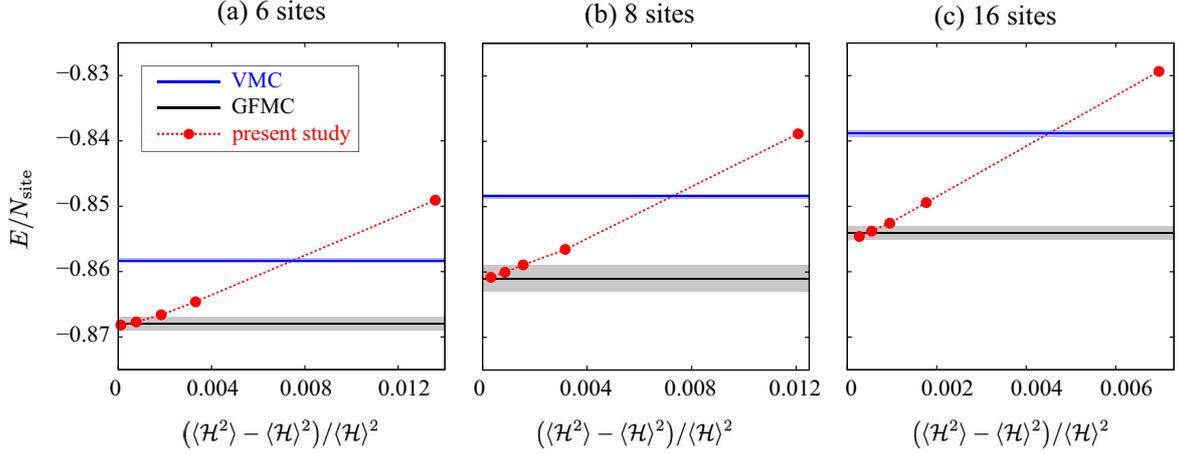}
\caption{
(Color online)
Energy for the one-dimensional spinless Holstein model [Eq.~(\ref{Eq_H_Holstein})] with $\omega/t = 1$ and $g/t = 1.5$ for 6-, 8-, and 16-site systems. 
The results of the present study are given by red dots. 
In each figure, from right to left, the hidden variable density $\alpha$ employed in the variational wave function varies as 0, 2, 4, 8, and 32. 
The horizontal axis is the energy variance $ \Delta_{\rm var } = ( \langle {\mathcal H } ^2 \rangle  -  \langle{\mathcal H }\rangle^2 ) / \langle {\mathcal H } \rangle^2$. 
For comparison, energies obtained by the variational Monte Carlo (VMC)~\cite{PhysRevB.89.195139} and Green's function Monte Carlo (GFMC) method~\cite{PhysRevB.53.9676} are shown, 
where the shaded regions show the size of error bar.  
}
\label{fig_1D_Holstein_ene}
\end{center}
\end{figure*}

Figure~\ref{fig_1D_Holstein_ene}  shows the results for the energy per site for different hidden variable densities $\alpha = 0$, 2, 4, 8, and 32.  
In each panel, the vertical axis is the energy per site and the horizontal axis is the energy variance 
$ \Delta_{\rm var } = ( \langle {\mathcal H } ^2 \rangle  -  \langle{\mathcal H }\rangle^2 ) / \langle {\mathcal H } \rangle^2$. 
The energy variance becomes zero if we obtain exact ground-state wave function (in more general, for any eigenstates of the Hamiltonian). 
Therefore, the energy variance tells us how close the variational wave function is to the ground state.   
As we can see, the accuracy becomes better with increasing $\alpha$, and in all the cases, the energy for $\alpha = 8$, 16, 32 agrees with those obtained by numerically exact Green's function Monte Carlo method~\cite{PhysRevB.53.9676} within one standard error.
In the case of the 16-site system, extrapolation to zero energy variance limit seems to overshoot the ground state energy estimated by the Monte Carlo method. 
However, the deviation is still within two standard errors. 
Note that our wave function method is variational; therefore, the energy never becomes lower than that of the ground state.

Compared to the result of the previous study (blue lines)~\cite{PhysRevB.89.195139}, our wave function improves the accuracy substantially. 
The improvement can be ascribed to the synergetic effect of the introduction of electron-phonon entangled wave function $ | \Psi_{\rm el \mathchar`- ph}  \rangle$ in Eq. (\ref{Eq_Psi_el-ph}) and the usage of the flexible RBM correlation factor in Eq. (\ref{Eq.form_RBM}). 
In Ref.~\citen{PhysRevB.89.195139}, the results for the wave function without the electron-phonon correlation factor $| \Psi_{\rm el } \rangle \otimes | \Psi_{\rm ph} \rangle$, in which the electron and phonon degrees of freedom are decoupled, are also shown;  
the obtained energies are  $-0.7749(2)$, $-0.7739(2)$, and $-0.7732(4)$ for 6-, 8-, and 16-site systems, respectively. 
By incorporating primitive correlations between electrons and phonons using $| \Psi_{\rm el \mathchar`- ph}  \rangle$, the results improve to $-0.84908(3)$, $-0.83884(3)$, $-0.82933(2)$ for for 6-, 8-, and 16-site systems, respectively ($\alpha=0$ result in Fig.~\ref{fig_1D_Holstein_ene}). 
The RBM correlation factor $ P^{\rm RBM}_{\rm el \mathchar`- ph} (\sigma)$ further improves the energy, achieving the agreement with the Monte Carlo results within one standard error.

\subsubsection{Charge structure factor} 
When the electron-phonon coupling becomes large, 
the ground state of the one-dimensional half-filled Holstein model  [Eq.~(\ref{Eq_H_Holstein})] changes from TLL to charge density wave (CDW) state. 
Here, we show the results for the charge structure factor for $(\omega/t = 0.1, \ g^2 / \omega^2 = 2)$ and $(\omega/t = 0.1, \ g^2 / \omega^2 = 20)$. 
The former gives the TLL ground state, and the latter gives rise to the CDW instability. 
In the present Hamiltonian, the charge-charge correlation function has a peak at the momentum $ q = \pi$, and the charge structure factor is given by 
\begin{eqnarray}
S_{\rm c} (\pi) =   \frac{1}{N_{\rm site}^2}  \sum_{i,j}   (-1)^j   \Bigl \langle   \bigl ( \hat{n}_i - \frac{1}{2} \bigr )   \bigl ( \hat{n}_{i+j} - \frac{1}{2}  \bigr) \Bigr \rangle. 
\end{eqnarray}

\begin{figure}[t]
\begin{center}
\includegraphics[width=0.45\textwidth]{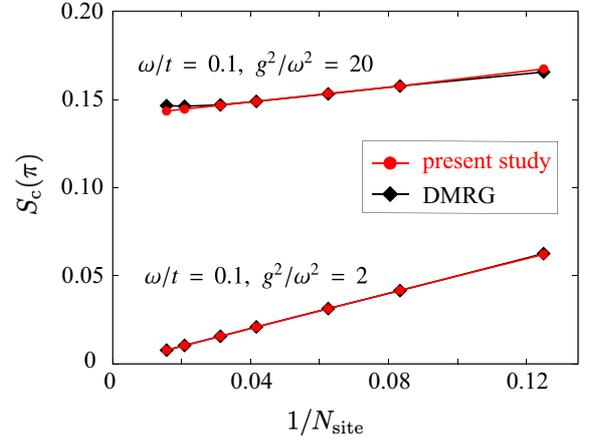}
\caption{
(Color online)
Results for charge structure factor $S_{\rm c} (\pi)$ for one-dimensional spinless Holstein model [Eq.~(\ref{Eq_H_Holstein})] with $(\omega/t = 0.1, \ g^2 / \omega^2 = 2)$ and $(\omega/t = 0.1, \ g^2 / \omega^2 = 20)$.
System sizes used in the calculations are 8, 12, 16, 24, 32, 48, and 64. 
Our results (red dots) are compared with those (black diamonds)~\cite{FEHSKE2005699} obtained by density-matrix renormalization group (DMRG). 
}
\label{fig_1D_Holstein_Sc}
\end{center}
\end{figure}

Before showing the results for $S_{\rm c} (\pi) $, we show the detail of the parameter setting.   
We employ the periodic boundary condition to compare our results with those of Ref. ~\citen{FEHSKE2005699}.
In the TLL phase with the particle-hole symmetry, we can impose the relationship in the $\{ c_{m_i}(n^{\rm s}_i) \}$ coefficients as described in the previous section: 
 $c_{2k} (n^{\rm s}_i = 1) =  c_{2k} (n^{\rm s}_i = -1)$ and $c_{2k + 1 } (n^{\rm s}_i = 1) = -  c_{2k + 1 } (n^{\rm s}_i = -1)$ with some non-negative integer $k$.
The number of independent phonon coefficients  $\{ c_{m_i}(n^{\rm s}_i) \}$  is  $m_{\rm max} \! +\! 1$. 
On the other hand, in the CDW phase, we have charge-rich and charge-poor sites, and the particle-hole symmetry at each site is broken. 
The actual ground state for the finite-size system is made of superposition of (rich, poor, rich, poor, ...) and (poor, rich, poor, rich, ...) CDW patterns, and the symmetry is recovered. 
However, in the present study, we assume one of two patterns. 
In this case, in the notation of $n^{\rm s}_i$, the CDW corresponds to the ferroic order favoring either $(1, 1, 1, 1,....)$ or $(-1, -1, -1, -1,...)$ patterns. 
Correspondingly, we prepare independent $c_{m_i}(n^{\rm s}_i )$ parameters for majority and minority $n^{\rm s}_i$ occupations. 
Thus, $c_{m_i} (n^{\rm s}_i \! = \! 1 )$ and $c_{m_i} (n^{\rm s}_i \! = \! -1 )$ become independent and  the number of independent phonon coefficients  $\{ c_{m_i}(n^{\rm s}_i) \}$  is  $2 (m_{\rm max} \! +\! 1)$.

In the TLL (CDW) phase for $\omega/t = 0.1$ and  $g^2 / \omega^2 = 2$  ($\omega/t = 0.1$ and $g^2 / \omega^2 = 20$), we set maximum phonon occupation number $m_{\rm max}$ at each site to be 7 (31),
and the number for visible units for phonon configuration becomes $M = 3$ ($M=5$) per site.
The total number of visible units is given by $N_{\rm visible} = N_{\rm site}(M+1)$. 
We confirm the convergence of results with respect to $m_{\rm max}$. 

As for the $f_{ij}$ parameters, we take 4-site unit-cell structure in common, which is large enough to study both TLL and commensurate CDW phases. 
The hidden variable density $\alpha$ is set to be $\alpha= 8$.

Figure~\ref{fig_1D_Holstein_Sc} shows the calculated charge structure factor $S_{\rm c} (\pi)$ for  $(\omega/t = 0.1, \ g^2 / \omega^2 = 2)$ and $(\omega/t = 0.1, \ g^2 / \omega^2 = 20)$. 
In the former case, $S_{\rm c} (\pi)$ vanishes as $N_{\rm site } \rightarrow \infty $, showing the absence of CDW order.
In the latter case, CDW order is confirmed by nonzero $S_{\rm c} (\pi)$  with $N_{\rm site} \rightarrow \infty $. 
We also compare our results with those~\cite{FEHSKE2005699} obtained by density-matrix renormalization group (DMRG)~\cite{PhysRevLett.69.2863}. 
They show good agreements, confirming the accuracy of our methods.

\subsection{Discussion}
In the present study, all the variational parameters are taken to be real. 
In this case, the RBM correlation factor $ P^{\rm RBM}_{\rm el \mathchar`- ph} (\sigma)$ always gives positive weight, i.e., it just controls the amplitude of the wave function and does not change the node of the wave function. 
The sign of the wave function is determined by electron-phonon entangled wave function $| \Psi_{\rm el \mathchar`- ph}  \rangle$. 
Therefore, we have paid close attention to the initial sign of the phonon coefficients $\{ c_{m_i}(n^{\rm s}_i) \}$ as described in the previous sections. 

In principle, if we introduce a complex RBM correlation factor, the node of the wave function can change. 
Even in that case, we believe that a good initial guess for the sign of phonon coefficients will make the optimization easier. 
Therefore, although it is straightforward to apply the machine learning wave functions to more general fermion-boson coupled Hamiltonians once the mapping between fermion-boson and visible-unit configurations is defined, 
it will be helpful to guess the sign structure of the exact wave functions (at least at primitive level) to obtain good accuracy. 

\section{Calculations of excited states} 
\label{sec_excited}

So far, almost all the variational studies using machine learning techniques have focused on approximating the ground state wave function. 
However, the calculations of excited states also give important information such as the ground state degeneracy, the size of the excitation gap, and low-lying dispersion of excitations. 

The first attempt to obtain excited states with machine-learning wave function has been performed in Ref.~\citen{PhysRevLett.121.167204}. 
In this study, we present another approach to obtain excited states. 
We apply the method to the one-dimensional Heisenberg model and show that our method gives better accuracy than that obtained in Ref.~\citen{PhysRevLett.121.167204} 
even though the present neural network is much more compact than that employed in Ref.~\citen{PhysRevLett.121.167204}. 
We also apply the method to the two-dimensional Heisenberg model with frustration and demonstrate good accuracy of the method.

\subsection{Method}

\subsubsection{General idea}
For finite size systems, eigenstates of many-body Hamiltonians with several symmetries are labelled by quantum numbers. 
Therefore, we can use quantum number projection to obtain excited states: 
We find the lowest energy eigenstates for different quantum number sectors; then we obtain excited states characterized by different quantum numbers than that of the ground state. 
This approach is taken commonly in the present study and Ref.~\citen{PhysRevLett.121.167204}.~\cite{note_excited} 
The difference in the schemes comes from the way of enforcing quantum numbers to the many-body wave function.

In the present study, we focus on translationally invariant systems, in which the total momentum ${\bf K}$ is a good quantum number. 
By applying the momentum projection, we can discuss the dispersion of the excited states.  
With a translation operator $T_{\bf R}$ shifting all the particles by the amount ${\bf R}$, 
the many-body wave function with the total momentum ${\bf K}$ is transformed as 
\begin{eqnarray}
\label{Eq_wf_sym}
 T_{\bf R}  \Psi_{\bf K}( {\bf r } ) \equiv \Psi_{\bf K}( {\bf r }+ {\bf R} ) = e^{i {\bf K} \cdot {\bf R} }  \Psi_{\bf K}( {\bf r } ), 
\end{eqnarray}  
where ${\bf r}$ denotes the real-space configuration of the particles ${\bf r} = ({\bf r}_1, {\bf r}_2, ..., {\bf r}_{N_{\rm particles} })$.
In the following, we discuss, in detail, how to make the wave function satisfy the symmetry in Eq.~(\ref{Eq_wf_sym}).


\subsubsection{Quantum number projection}
Here, we discuss how to apply the momentum number projection to the wave function. 
For simplicity, let us give an explanation using the one-dimensional $S \! =\! \frac{1}{2}$ spin models with the periodic boundary condition.
The spin configuration is specified by $\sigma = (\sigma_1, \sigma_2, ..., \sigma_{N_{\rm site}})$ with $\sigma_i = 2 S^z_i = \pm1$. 
$S^z_i$ is the $z$-component of $S \! =\! \frac{1}{2}$ spin at site $i$.

\paragraph{Method in Ref.~\citen{PhysRevLett.121.167204}.}
For a spin configuration, we can generate shifted configurations by applying the translation operators. 
Among the generated spin configurations including the original configuration, we choose a canonical configuration $\sigma_{\rm canonical}$. 
In Ref.~\citen{PhysRevLett.121.167204}, the canonical configuration  $\sigma_{\rm canonical}$ is chosen to be the lexicographically smallest one.
By introducing an operator $T$ shifting spin configurations by one site,
the amplitude of the wave function for a configuration  $T^n  \sigma_{\rm canonical} $ ($ 0 \! \leq  \! n  \! < \!  N_{\rm site}$)  is given by 
\begin{eqnarray}
 \Psi_K ( T^n  \sigma_{\rm canonical} ) = e^{i n K}   \Psi ( \sigma_{\rm canonical} )  .
 \end{eqnarray}  
Here, instead of calculating the amplitude for  $T^n  \sigma_{\rm canonical} $  directly, the amplitude is given by referring to that of the canonical configuration $\sigma_{\rm canonical}$. 
With this, the wave function on the left-hand side satisfies the proper symmetry in Eq. (\ref{Eq_wf_sym})
even when we do not impose any translation symmetry constraint on the wave function on the right-hand side $  \Psi ( \sigma_{\rm canonical} )  $. 
In Ref.~\citen{PhysRevLett.121.167204}, $ \Psi ( \sigma_{\rm canonical} ) $ is prepared by the RBM or three-layer feedforward neural networks (FFNN).

\paragraph{Present Scheme.}
In the present study, we employ the scheme in Refs.~\citen{Mizusaki} and \citen{doi:10.1143/JPSJ.77.114701}.
The wave function projected onto the total momentum $K$ sector is given by 
\begin{eqnarray}
 \Psi_K ( \sigma ) =  \sum_{n = 0}^{N_{\rm site} -1 } e^{ - i n K}   \Psi ( T^n   \sigma ). 
 \label{Eq.momentum_proj}
 \end{eqnarray}
Is it easy to show that the wave function on the left-hand side is transformed according to Eq. (\ref{Eq_wf_sym}). 
Note again that the wave function on the right-hand side does not necessarily satisfy the symmetry in Eq. (\ref{Eq_wf_sym}). 
To represent $\Psi(\sigma)$, we employ the RBM wave function with $N_{\rm visible} = N_{\rm site}$, which reads
\begin{eqnarray}
 \Psi(\sigma) \! \! \! \!  &=& \! \! \!  \!   \sum_{ \{ h_k\}}  \exp \left (   \sum_{i,k} W_{ik}  \sigma_i h_k +\sum_{k} b_k h_k \right ) \nonumber \\ 
 &=& \! \! \! \!  \prod_k   2 \cosh \left( b_k + \sum_{i} W_{ik}  \sigma_i  \right ).  
\end{eqnarray}
$b_k$ is the bias on the hidden units, and $W_{ik} $ is the interaction between the visible and hidden units. 
As in Sec.~\ref{Sec:el_ph}, we neglect the bias term for the visible units. 
In order to make it possible to represent the sign change of the wave function,
{\it we take the $b_k$ and $W_{ik} $ parameters to be complex variables. }

\paragraph{Comparison between the present scheme and that in Ref.~\citen{PhysRevLett.121.167204}.} 

Let us show the difference in the above two schemes using the simple one-dimensional four-site spin model. 
In this case, for example, the following four spin configurations are related with each other by the translation operator:  
\begin{eqnarray}
\sigma_0 &=& ( \uparrow,  \uparrow, \downarrow, \downarrow )  \nonumber \\ 
\sigma_1 &=& ( \downarrow, \uparrow,  \uparrow,  \downarrow ) = T \sigma_0 \nonumber \\ 
\sigma_2 &=& ( \downarrow , \downarrow, \uparrow,  \uparrow  ) = T^2 \sigma_0 \nonumber  \\
\sigma_3 &=& (  \uparrow ,  \downarrow,\downarrow ,   \uparrow ) = T^3 \sigma_0
\end{eqnarray} 
Let us take $\sigma_{\rm canonical }$ to be $\sigma_0$. 
Then, the wave function in Ref.~\citen{PhysRevLett.121.167204} is given by 
\begin{eqnarray}
 \Psi_K ( \sigma_0 ) &=&    \Psi ( \sigma_0 )  \nonumber \\ 
 \Psi_K ( \sigma_1 ) &= & e^{iK}   \Psi ( \sigma_0 )  \nonumber \\ 
  \Psi_K ( \sigma_2 ) &= & e^{2 iK}   \Psi ( \sigma_0 )  \nonumber \\ 
   \Psi_K ( \sigma_3 ) &= & e^{3 iK}    \Psi ( \sigma_0 )   \nonumber 
 \end{eqnarray}  
On the other hand, the wave function in the present scheme reads  
\begin{eqnarray}
 \Psi_K ( \sigma_0 )  \! \! \! \!  &=&  \! \! \! \!   \Psi ( \sigma_0 )  + e^{- iK}  \Psi ( \sigma_1 ) + e^{-2 iK}  \Psi ( \sigma_2 ) + e^{-3 iK}  \Psi ( \sigma_3 )
 \nonumber \\ 
 \Psi_K ( \sigma_1 )   \! \! \! \!  &=&   \! \! \! \!   \Psi ( \sigma_1 )  + e^{- iK}  \Psi ( \sigma_2 ) + e^{-2 iK}  \Psi ( \sigma_3 ) + e^{-3iK}  \Psi ( \sigma_0 )
 \nonumber \\ 
 \Psi_K ( \sigma_2 )   \! \! \! \!  &=&   \! \! \! \!   \Psi ( \sigma_2 )  + e^{- iK}  \Psi ( \sigma_3 ) + e^{-2 iK}  \Psi ( \sigma_0 ) + e^{-3 iK}  \Psi ( \sigma_1 )
 \nonumber \\ 
 \Psi_K ( \sigma_3 )  \! \! \! \!  &=&  \! \! \! \!    \Psi ( \sigma_3 )  + e^{- iK}  \Psi ( \sigma_0 ) + e^{-2 iK}  \Psi ( \sigma_1 ) + e^{-3 iK}  \Psi ( \sigma_2 ).
 \nonumber 
 \end{eqnarray}

When we define the hidden variable density $\alpha$ as $\alpha = N_{\rm hidden} / N_{\rm visible}$ ($= N_{\rm hidden} / N_{\rm site}$) as in Sec.~\ref{Sec:el_ph}, 
the computational cost of the scheme in Ref.~\citen{PhysRevLett.121.167204} scales as ${\mathcal O} ( \alpha N_{\rm site} ^2)$. 
On the other hand, the present scheme scales as ${\mathcal O} ( \alpha N_{\rm site} ^3)$ because we need to compute the summation over $n$ in Eq.~(\ref{Eq.momentum_proj}), which gives an additional factor of $N_{\rm site}$.
Therefore, when the hidden variable density $\alpha$ is the same, the scheme in Ref.~\citen{PhysRevLett.121.167204}  is computationally cheaper. 
However, as we show in the next section, the present scheme is much more accurate with the same $\alpha$.
In the scheme in Ref.~\citen{PhysRevLett.121.167204},  if the value of $\alpha$ giving the same accuracy as that in the present scheme becomes comparable to $N_{\rm site}$, 
the computational cost becomes comparable between the two methods. 

Furthermore, when the number of variational parameters becomes large, numerical optimization becomes difficult. 
Therefore, in general, it helps to save the number of variational parameters for achieving stable optimizations. 
In this sense, the present scheme has an advantage because we can achieve much better accuracy with smaller $\alpha$ and hence a smaller number of variational parameters.

\subsection{Results}

\subsubsection{One-dimensional antiferromagnetic Heisenberg model}

\begin{figure}[t]
\begin{center}
\includegraphics[width=0.48\textwidth]{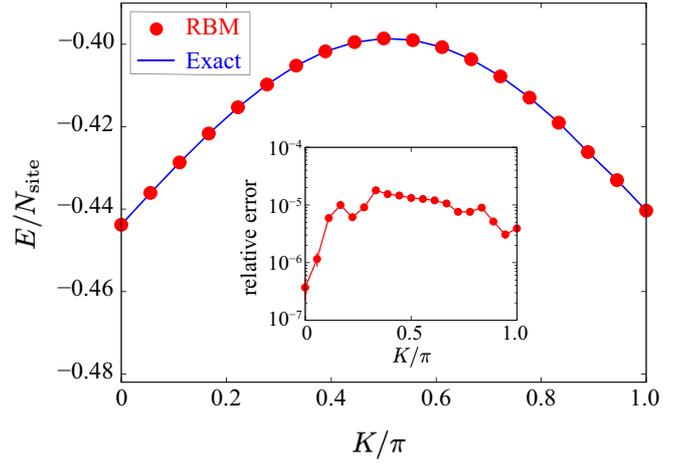}
\caption{
(Color online)
Energy per site as a function of total momentum $K$ for the one-dimensional antiferromagnetic Heisenberg model.  
Red dots show the RBM results with the hidden variable density $\alpha=1$.
The size of error bars is smaller than the symbol size. 
The exact results obtained by the exact diagonalization are shown by the blue curve. 
The inset shows the relative error of the energy $(E_{\rm RBM} - E_{\rm exact}) / E_{\rm GS}$ with the RBM energy $E_{\rm RBM}$, the exact energy $E_{\rm exact}$, and the exact ground state energy $E_{\rm GS}$. }
\label{fig_1D_Heis_ene}
\end{center}
\end{figure}

First, we show the result of $S=\frac{1}{2}$ antiferromagnetic Heisenberg model on the one-dimensional spin chain. 
The Hamiltonian is given by 
\begin{eqnarray} 
 {\mathcal H} =  J \sum_{ i=1 } ^{ N_{\rm site}}  {\bf S}_i  \cdot {\bf S}_{i+1}.
 \end{eqnarray}
Here  ${\bf S}_i$ is the spin-$1/2$ operator at site $i$ and $J$ is the Heisenberg exchange interaction. 
We take $J$ as energy unit, i.e., we set $J=1$. 
As in Ref.~\citen{PhysRevLett.121.167204}, we take 36 site systems ($N_{\rm site} = 36$) and impose periodic boundary condition. 
The relation between the visible units and physical spins is given by $\sigma_i = 2 S^z_i$, as we described above. 
The number of visible units is $N_{\rm visible} = N_{\rm site}$.  
We do not impose any symmetry in $b_k$ or $W_{ik}$, and the numbers of independent $b_k$ and $W_{ik}$ parameters are $\alpha N_{\rm visible} $ and $\alpha N_{\rm visible}^2$, respectively. 
The total number of variational parameters amounts to $2 \alpha N_{\rm visible}  (N_{\rm visible} + 1)$, 
where the factor of 2 comes from the fact that $b_k$ and $W_{ik}$ parameters have both real and imaginary parts. 
The optimization of the variational parameters is done in the very same way as in Sec.~\ref{Sec:el_ph}, i.e., we use the SR method. 

We optimize the RBM wave function with the hidden variable density $\alpha=1$ for each momentum $K$ sector and compare the energy with the exact results (Fig.~\ref{fig_1D_Heis_ene}). 
We find that the relative error of the RBM energy is at most $2\times 10^{-5}$, showing an excellent agreement with the exact energy. 
Indeed, the present RBM energy is much more accurate than that obtained in Ref.~\citen{PhysRevLett.121.167204}: 
In Ref.~\citen{PhysRevLett.121.167204}, the results obtained by the RBM with $\alpha=3$ show the relative error of about $7 \times10^{-3}$ at the momentum in which the accuracy is the worst (around $K= \frac{2}{3} \pi$).
The best accuracy is achieved at $K=0$ in both cases, the relative error is about $4 \times10^{-7}$ (the present study) and  $2 \times10^{-5}$ (Ref.~\citen{PhysRevLett.121.167204}).

As we already described, the way of imposing total momentum is different in the two methods, which leads to a significant difference in the accuracy. 
The present RBM wave function gives better accuracy even though the number of hidden units is smaller. 
To improve the accuracy, in Ref.~\citen{PhysRevLett.121.167204}, the three-layer FFNN is introduced. 
The size of the FFNN is characterized by the first hidden-layer variable density $\alpha_1 = 2$ and the second hidden-layer variable density $\alpha_2 = 0.5$. 
However, the accuracy of the FFNN is still worse than that obtained by the present RBM: 
the best accuracy is achieved at $K=0$ and the relative error is about $3 \times10^{-5}$, 
whereas the worst accurate momentum is at $K = \pi/ 2$ and the relative error is about $5 \times10^{-4}$. 



\subsubsection{Two-dimensional antiferromagnetic $J_1$-$J_2$ Heisenberg model}
We also apply the method to the $S=\frac{1}{2}$ antiferromagnetic $J_1$-$J_2$ Heisenberg model on the square lattice. 
The Hamiltonian is given by 
\begin{eqnarray} 
 {\mathcal H} =  J_1 \sum_{ \langle i, j \rangle}  {\bf S}_i  \cdot {\bf S}_j
 +  J_2 \sum_{ \langle \langle  i, j \rangle \rangle}  {\bf S}_i  \cdot {\bf S}_j, 
 \end{eqnarray}
where ${\bf S}_i$ is the spin-$1/2$ operator at site $i$, and $J_1$ ($J_2$) is the nearest neighbor (next nearest neighbor) exchange interaction. 
We take $J_1$ as energy unit, i.e., we set $J_1=1$. 
$ \langle i, j \rangle$ and $\langle \langle i, j \rangle \rangle$ denote pairs of nearest neighbor and next nearest-neighbor sites, respectively. 
In the $S=\frac{1}{2}$ antiferromagnetic $J_1$-$J_2$ Heisenberg model on the square lattice, 
the next nearest-neighbor $J_2$ interaction gives frustration and the model may host spin liquid ground state around $J_2 / J_1$ = 0.5.
However, it is still an open problem whether the spin liquid ground state is realized or not. 
For the frustrated systems, whereas the quantum Monte Carlo methods suffer from the negative sign problem, 
the RBM scheme is free from the sign problem and can be applied to this challenging problem.

\begin{figure}[t]
\begin{center}
\includegraphics[width=0.45\textwidth]{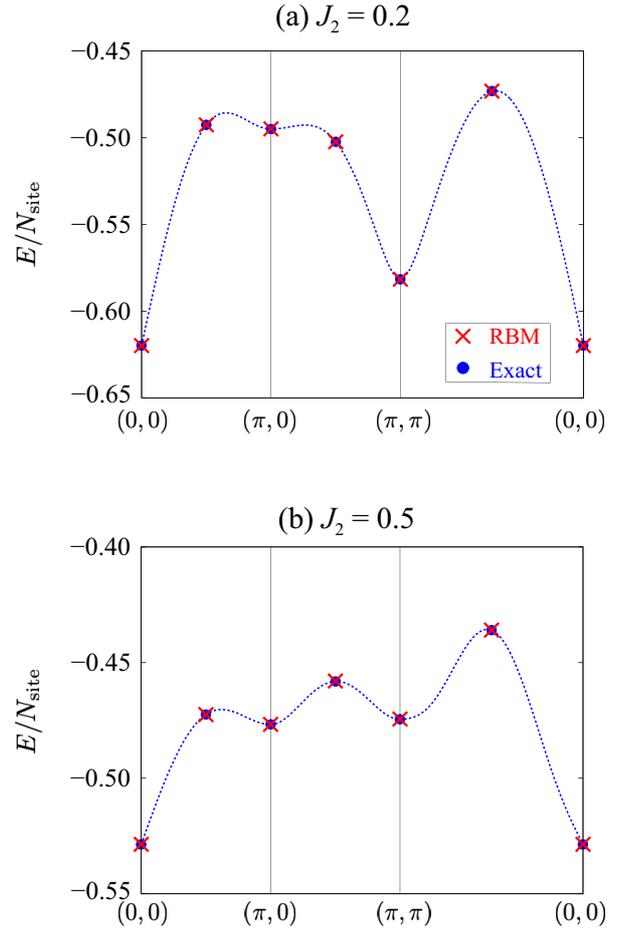}
\caption{
(Color online)
Energy per site as a function of total momentum ${\bf K}$ for the two-dimensional $J_1$-$J_2$ Heisenberg model with (a) $J_2 = 0.2$ and (b) $J_2 = 0.5$. 
The RBM results with the hidden variable density of $\alpha=2.5$ are shown by red crosses. 
Blue dots show the exact results obtained by the exact diagonalization.
Blue dotted lines are just guide to the eye.  
}
\label{fig_2D_J1J2_ene}
\end{center}
\end{figure}

Here, as a benchmark, we employ the $4 \times 4$ lattice with the periodic boundary condition. 
In the case of the $4 \times 4$ lattice, there are six irreducible momenta 
$(0,0)$, $(\pi/2,0)$, $(\pi,0)$, $(\pi,\pi/2)$, $(\pi,\pi)$ and $(\pi/2,\pi/2)$.
For each momentum, we compute the lowest energy state with the RBM ($\alpha=2.5$) and compare them with the results obtained by the exact diagonalization. 
Fig.~\ref{fig_2D_J1J2_ene}(a) and (b) show the results for $J_2/J_1 = 0.2$ and 0.5, respectively. 
As can be seen, the agreement between the RBM and the exact results is excellent. 
The relative errors are at most $1 \times 10 ^{-4}$ and  $4 \times 10 ^{-4}$ for $J_2/J_1 = 0.2$ and 0.5, respectively.

\subsection{Discussion}

We have shown that the complex RBM can well describe the excited states of the quantum spin Hamiltonians.
The method seems to work well even in the frustrated spin systems, in which the node structure of the wave function becomes crucial. 
Therefore, it opens a way to investigate, for example, the existence of the spin gap in the frustrated systems. 

For large system sizes, the number of variational parameters would become large to achieve high accuracy. 
Because it is difficult to perform stable optimization of a large number of parameters, it is helpful to reduce the number of parameters. 
As is discussed in Ref.~\citen{PhysRevB.96.205152}, by combining the RBM with other powerful wave functions, the same accuracy would be achieved by smaller number of variational parameters.  
It is an interesting future issue to implement such a combination to study excited states. 

\section{Summary} 
\label{sec_summary}

We have discussed two extensions of the machine learning variational method. 
The first extension is the application to the fermion-boson coupled Hamiltonians. 
We have studied the Holstein model as a representative and showed that the present RBM-based variational wave function outperforms the previous variational wave function.   
The improvement is achieved by preparing electron-phonon-entangled wave function and improving the electron-phonon correlation factor using the RBM.   

The second extension is the calculation of excited states. 
This is achieved by imposing quantum numbers to the wave functions.
The difference between the present method and that in Ref.~\citen{PhysRevLett.121.167204} lies in the way of imposing quantum numbers.
By applying it to the one-dimensional $S=1/2$ Heisenberg model, we have shown that the present scheme gives better accuracy than that achieved in Ref.~\citen{PhysRevLett.121.167204}, 
even though the present neural network is more compact. 
We have also performed the benchmark using the two-dimensional $S=1/2$ $J_1$-$J_2$ Heisenberg model, and showed an excellent agreement with the exact results.  

Finally, we briefly discuss several future directions. 
As for the first part, an application to more general fermion boson systems would be of great interest.  
It will open a way to study real materials, which sometimes have complicated forms of interactions. 
As for the second part, the information of excited states is crucial to reveal the nature of quantum spin liquids in the frustrated spin systems, if the spin liquid exists as a stable phase. 
For example, the two-dimensional $S=1/2$ $J_1$-$J_2$ Heisenberg model, which we have studied in this paper, is one of the candidate Hamiltonians to host quantum spin liquid as a ground state. 
It is intriguing to perform calculations for larger system sizes and investigate the nature of the quantum spin liquid.   
We finally note that the present method to calculate excited states can be easily generalized to, e.g., fermion and fermion-boson coupled systems, which is also left as an important future issue.

\begin{acknowledgment}


{\it Acknowledgments.}
We are grateful for fruitful discussions with Kenny Choo, Giuseppe Carleo, Takahiro Ohgoe, Masatoshi Imada, and Youhei Yamaji. 
In particular, we thank Kenny Choo and Takahiro Ohgoe for providing us the raw data in Refs.~\citen{PhysRevLett.121.167204} and \citen{PhysRevB.89.195139}, respectively. 
The implementation of the machine-learning method is done based on the mVMC package~\cite{MISAWA2019447}.
This work was supported by Grant-in-Aids for Scientific Research (JSPS KAKENHI) (Grants No. 16H06345, No. 17K14336 and No. 18H01158).

\end{acknowledgment}

\bibliographystyle{jpsj}
\bibliography{main}

\begin{thebibliography}{10}
\bibitem{note_SR}
The equivalent optimization scheme is called natural gradient in the artificial
  intelligence community~\cite{IEEE_Amari,neural_Amari}.
\bibitem{note_excited}
In Ref.~\citen{PhysRevLett.121.167204}, independently from the quantum number
  approach, excited states are also constructed by orthogonalizing the wave
  function with respect to the variational ground state wave function. In the
  present study, we do not discuss such kind of approach.
\end{thebibliography}


\end{document}